\documentclass[aps,pre,10pt,twocolumn,superscriptaddress]{revtex4-2}
\usepackage{graphicx}
\usepackage{epstopdf}
\usepackage{xcolor}
\usepackage{amsmath}
\usepackage{amssymb}
\usepackage[utf8]{inputenc}
\usepackage{graphicx,float,hyperref}
\usepackage{comment}
\usepackage{natbib}
\usepackage{graphicx,latexsym} 

\definecolor{blueOA}{rgb}{0.1, 0.0, 0.91}
\newcommand{\OA}[1]{\textcolor{blueOA}{#1}}

\newcommand{\be}{\begin{equation}}
\newcommand{\bw}{\begin{widetext}}
\newcommand{\ew}{\end{widetext}}
\newcommand{\bea}{\begin{eqnarray}}
\newcommand{\bc}{\begin{center}}            
\newcommand{\ee}{\end{equation}}
\newcommand{\eea}{\end{eqnarray}}
\newcommand{\ec}{\end{center}}

\begin{document}

% Use the \preprint command to place your local institutional report number 
% on the title page in preprint mode.
% Multiple \preprint commands are allowed.
%\preprint{}

% \title{Asymmetric harmonic quantum Otto engine: Frictional effects, performance bounds and operational modes} 
\title{The asymmetric Otto engine: frictional effects on performance bounds and operational modes} 

\author{Varinder Singh}
\email{varinder@ibs.re.kr}
\affiliation{Center for Theoretical Physics of Complex Systems, Institute for Basic Science (IBS), Daejeon 34126, Korea}
%\affiliation{Department of Physics, Ko\c{c} University, 34450 Sar\i{}yer, Istanbul, Turkey}
%
\author{Vahid Shaghaghi}
%\email{vshaghaghi@uninsubria.it}
\affiliation{Center for Theoretical Physics of Complex Systems, Institute for Basic Science (IBS), Daejeon 34126, Korea}
\affiliation{Center for Nonlinear and Complex Systems, Dipartimento di Scienza e Alta
Tecnologia, Universit\`{a} degli Studi dell’Insubria, via Valleggio 11, 22100 Como, Italy}
\affiliation{
Istituto Nazionale di Fisica Nucleare, Sezione di Milano, via Celoria 16, 20133 Milano, Italy}

\author{Tanmoy Pandit}
%\email{tanmoy.pandit@mail.huji.ac.il}
\affiliation{Fritz Haber Research Center for Molecular Dynamics, Hebrew University of Jerusalem, Jerusalem 9190401, Israel}

\author{Cameron Beetar}
\affiliation{The Laboratory for Quantum Gravity \& Strings, Department of Mathematics \& Applied Mathematics, University of Cape Town, Rondebosch, Cape Town, South Africa}

\author{Giuliano Benenti}
%\email{vshaghaghi@uninsubria.it}
\affiliation{Center for Nonlinear and Complex Systems, Dipartimento di Scienza e Alta
Tecnologia, Universit\`{a} degli Studi dell’Insubria, via Valleggio 11, 22100 Como, Italy}
\affiliation{Istituto Nazionale di Fisica Nucleare, Sezione di Milano, via Celoria 16, 20133 Milano, Italy}
\affiliation{NEST, Istituto Nanoscienze-CNR, P.zza San Silvestro 12, I-56127 Pisa, Italy}
\author{Dario Rosa}
\email{dario\_rosa@ibs.re.kr}
\affiliation{Center for Theoretical Physics of Complex Systems, Institute for Basic Science (IBS), Daejeon 34126, Korea}
\affiliation{Basic Science Program, Korea University of Science and Technology (UST), Daejeon - 34113, Korea}
%
 %\date{\today}

%\listoffigures

\begin{abstract}
We present a detailed study of an asymmetrically driven quantum Otto engine with a time-dependent harmonic oscillator as its working medium.  We obtain analytic expressions for the upper bounds on the efficiency of the engine for two different driving schemes having asymmetry in the expansion and compression work strokes. We show that the Otto cycle under consideration cannot operate as a heat engine in the low-temperature regime. Then, we   show that the friction in the expansion stroke is significantly more detrimental to the performance of the engine as compared to the friction in the compression stroke. Further, by comparing the performance of the engine with sudden expansion, sudden compression, and both sudden strokes, we uncover a pattern of connections between the operational points, and we indicate the optimal operation regime for each case. Finally, we analytically characterize the complete phase diagram of the Otto cycle for both driving schemes and highlight the different operational modes of the cycle as a heat engine, refrigerator, accelerator, and heater. 
\end{abstract}

% insert suggested PACS numbers in braces on next line

\maketitle %\maketitle must follow title, authors, abstract and \pacs

\section{Introduction}
The study of thermal machines is a central topic in thermodynamics. In fact, the desire to improve upon the efficiency of steam engines led Carnot to discover the concept of Carnot efficiency, which is one of the cornerstones of thermodynamics  \cite{DilipBook}. The Carnot efficiency, $\eta_C=1-T_c/T_h$, provides an upper bound on the efficiency of all macroscopic heat engines working between two thermal reservoirs at temperatures $T_c$ and $T_h$ ($T_c<T_h$) \cite{Callenbook1985,DilipBook}. However, the reversible nature of the Carnot cycle makes it an impractical device since it requires an infinite amount of time to complete one cycle and therefore its power output vanishes. The quest for engines operating in finite time and with non-zero power output gave rise to the field of finite-time thermodynamics \cite{Salamon2001,Andresen2011,Berry1984,devosbook}. These practical engines, operating in finite time, undergo irreversible thermodynamic transformations and are subjected to frictional effects \cite{Salamon1981,CA1975,Rubin1979,Kosloff1984,Alicki1979}.
These effects make the engines less efficient and, consequently, the Carnot efficiency may not be reachable – even in the limiting case of vanishing power output. Such an observation suggests that, in the presence of frictional effects, it may be possible to derive bounds on the efficiency which are tighter than the usual Carnot efficiency bound \cite{Mohanta2023,Shiraishi2017,Feldmann2003,Peterson2019PRL}.

However, except for a few studies \cite{Feldmann2003,VOzgur2020}, most recent work concentrates on the general performance characteristics of the engines\cite{VS2022,Feldmann2000,Feldmann2006,Abah2016EPL,Deffner2020,Rezek2006,Rezek2017,Denzler2021,Cleverson2022,Solfanelli2020,Ozgur2017,Lee2021,RazzoliCarrega2023,Cavaliere2023,Carrega2023,Pritam2020,GeorgeFriction,VJ2018,FrancescoCarrega2019,Kiran2021,VarinderJohal,GeorgeSir2011}, rather than deriving explicit upper bounds on the efficiency (and comparing this to their friction-less counterparts). 
In this paper, we will continue the program initiated in \cite{VOzgur2020} to quantitatively estimate the effects of friction on the performance bounds of a quantum engine.
In detail, we will consider a quantum Otto engine having a time-dependent harmonic oscillator as its working medium \cite{Kieu2004,Rezek2006,VOzgur2020,Vahid2021,Assis2019,Assis2020A,Tanmoy2021}, and we will consider one of the two work strokes to be \textit{sudden} in time (while the other remains adiabatic), thus making the resulting cycle \textit{asymmetric} in the compression/expansion strokes.
The asymmetric nature of the Otto cycle breaks the time-reversal symmetry between the expansion and compression strokes and as a result time forward and reversed cycles are distinct \cite{Mohanta2023}. 
For our purposes, we will show analytically that the bounds on the efficiency of the two resulting configurations are \textit{qualitatively different}, with the efficiency of the engine working with a sudden compression being much higher than the engine working with a sudden expansion \footnote{We observe that recently, it was shown that the nonadiabatic nature of the driven compression/expansion strokes, incorporating friction effects, will reduce the efficiency well below the Carnot efficiency \cite{Peterson2019PRL,Shiraishi2017}. However, an analytic expression for the upper bound has, as yet, not been derived in terms of simple system and/or bath parameters.}.
% We will provide a heuristic explanation for this property.

In addition, we also discuss the effects of friction in the context of the different operational modes of the quantum harmonic Otto cycle; such as a heat engine, refrigerator, heater, and accelerator \cite{Solfanelli2020,Mohsen2023,Ishizaki2023,Cleverson2022,Leitch2022,ChenPoletti2021}. Our analysis will show that, in the presence of the aforementioned asymmetry, the Otto cycle does not perform as an engine or a refrigerator only, but that the operational, potentially useful~\cite{Piccione2021PRA,Dieguez2023}, modes of heater and accelerator become available, leading to a richer phase diagram that we explicitly characterize. 

Recently, the optimization and study of time asymmetric heat engines has attracted a considerable amount of attention among researchers \cite{Mohanta2023,Shastri2023,Zheng2016,Gingrich2014,Pal2016,Rana2014}. Our results complement these studies, providing a setup in which the effects of the asymmetry can be studied analytically and explicitly.

The rest of the paper is organized as follows. In Sec.~\ref{sec2}, we give a brief description of the model of a harmonic quantum Otto cycle.  Sec.~\ref{sec3} is devoted to the study of the asymmetric Otto cycle with a sudden expansion stroke and a quasistatic compression stroke, from which analytic expressions for the upper bounds on the efficiency and efficiency at maximum work are obtained. In Sec.~\ref{sec4}, we repeat the same analysis for the asymmetric Otto cycle with a sudden compression stroke and a quasi-static expansion stroke. In Sec.~\ref{sec5}, we compare the performance of the engine under these two different schemes and include comparisons of their performance with that of the scenario in which both expansion and compression strokes are sudden, and thus nonadiabatic in nature. In Sec.~\ref{sec6}, we plot a complete phase diagram for the Otto cycle and mark different operational modes of the harmonic Otto cycle in terms of system-bath parameters. In Sec.~\ref{sec7} we conclude the paper.

\section{Quantum Otto cycle}\label{sec2}

\begin{figure}
\centering
\includegraphics[width=8.6cm]{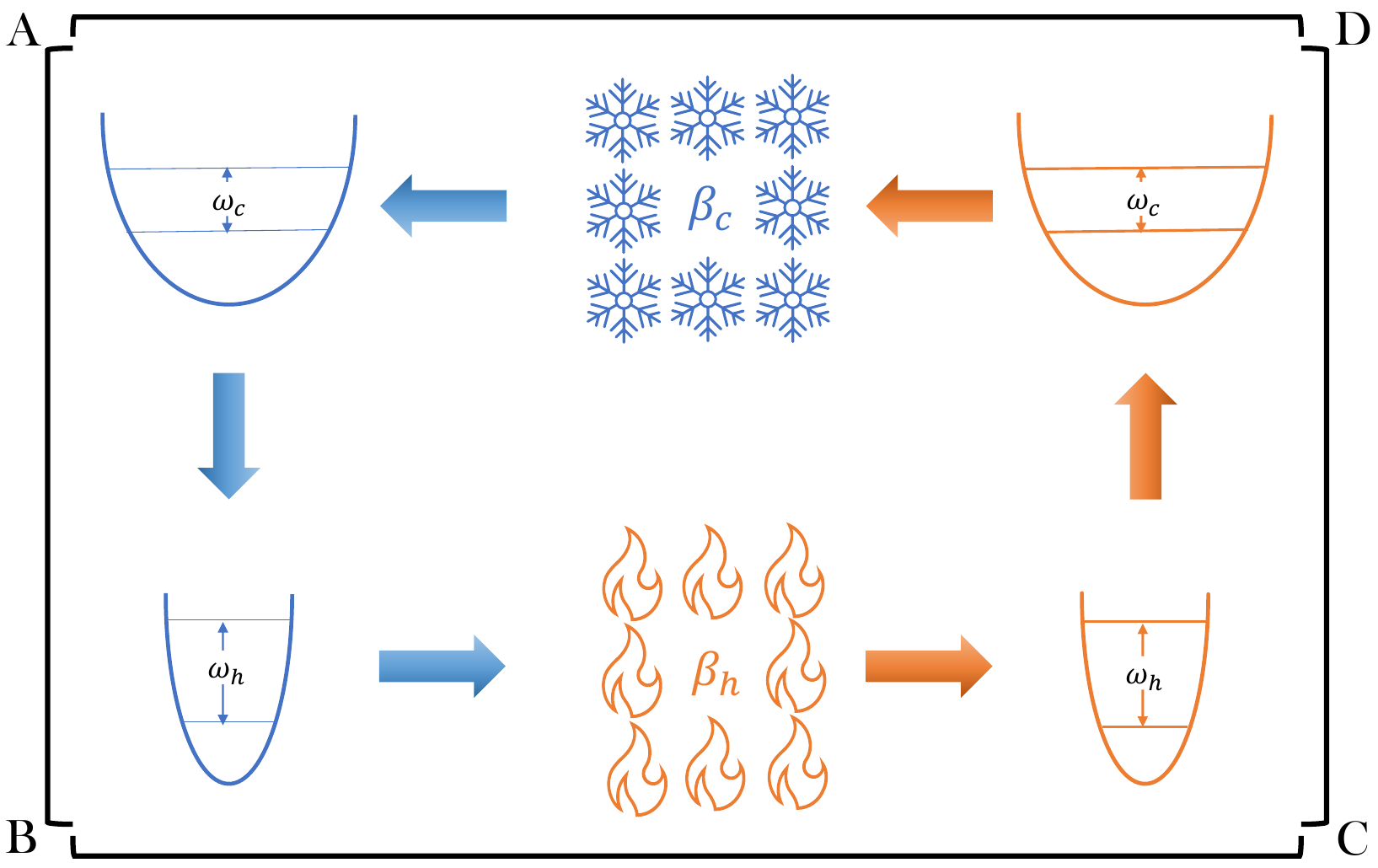}  
 % entropy.pdf: 0x0 px, 0dpi, nanxnan cm, bb=
\caption{Pictorial depiction of the Otto cycle. The thermodynamic cycle consists of four stages: two work (A$\rightarrow$ B and C$\rightarrow$ D) steps and two isochoric heat (B$\rightarrow$ C and D $\rightarrow$ A) steps.} \label{fig:cycle}
\end{figure}

The quantum Otto cycle is characterized by $4$ thermodynamic cyclic processes -- $2$ work strokes \footnote{These strokes are often called \textit{adiabatic} in the literature, with adiabatic referring to the term in its \textit{thermodynamic} sense, \textit{i.e.} with the system being isolated from its environment. However, in this paper, we will use the word adiabatic to indicate a process that is adiabatic in the \textit{quantum} sense, \textit{i.e.} which is quasi-static in nature and does not induce frictional effects.} and $2$ isochoric heat strokes -- which are schematically represented in Fig.~\ref{fig:cycle}. More precisely, these steps occur in the following order \cite{Lutz2012,Abah2016EPL}:

\begin{enumerate}
    \item Compression stroke $A\longrightarrow B$:  We assume that the system (a quantum harmonic oscillator, in this case) is initially in a thermal state characterized by the inverse temperature $\beta_c$ and frequency $\omega_c$. Then, while keeping the system isolated from its surroundings, the frequency is brought to the new value $\omega_h$ (with $\omega_h > \omega_c$) by means of an external driving protocol (which, in the adiabatic case, would be quasi-static). During this stroke, the energy of the harmonic oscillator increases, and therefore work is done on the system. On the other hand, the process is unitary and the von Neumann entropy of the system does not change.
    \item Hot isochore $B\longrightarrow C$: The harmonic oscillator is put in contact with a hot bath, having inverse temperature $\beta_h$ (with $\beta_h < \beta_c$). The frequency of the harmonic oscillator is kept constant and the system is allowed to thermalize with the hot bath, thus reaching the final inverse temperature $\beta_h$ \cite{CampoGoold} \footnote{In this work,  the working fluid is assumed to be fully thermalized in a finite time, which is a justifiable assumption provided that the heat transport coefficients are large. More generally,  the evolution of a quantum system in contact with a thermal bath is described by the Lindblad master equation. For the quantum harmonic Otto engine, details can be found in \cite{Rezek2006}. }.
    \item  Expansion stroke $C \longrightarrow D$: The system is again isolated from its surroundings, and its frequency is reduced to the initial value $\omega_c$ by means of a unitary driving protocol. During this step, the harmonic oscillator performs work.
    \item Cold isochore $D\longrightarrow A$: To close the cycle and bring the harmonic oscillator back to its initial setup, a cold bath (having inverse temperature $\beta_c$) is put in contact with the system. The harmonic oscillator is left to thermalize while keeping its frequency equal to $\omega_c$.    
\end{enumerate}

 The average  energies $\langle H \rangle$ of the oscillator during the four stages of the cycle are (in units of $\hbar\!=1\!$ and $k_B = 1$) \cite{Lutz2012,Abah2016EPL} 
\begin{equation}
\langle H\rangle_A =\frac{ \omega_c}{2}\text{coth}\Big(\frac{\beta_c  \omega_c}{2}\Big) ,
\end{equation}
\begin{equation}
\langle H\rangle_B =\frac{ \omega_h}{2} \lambda_{AB} \text{coth}\Big(\frac{\beta_c  \omega_c}{2}\Big) ,
\end{equation}
\begin{equation}
\langle H\rangle_C =\frac{ \omega_h}{2} \text{coth}\Big(\frac{\beta_h   \omega_h}{2}\Big) ,
\end{equation}
\begin{equation}
\langle H\rangle_D =\frac{ \omega_c}{2} \lambda_{CD} \text{coth}\Big(\frac{\beta_h  \omega_h}{2}\Big) ,
\end{equation}
%we have set $\hbar\!=\!k_\mathrm{B}=1$ for simplicity.
where the parameter $\lambda$ (either $\lambda_{AB}$ or $\lambda_{CD}$) is called the \textit{adiabaticity parameter} of the dynamics. It is a dimensionless quantity that depends on the nature of the frequency modulation, see Refs. \cite{Deffner2008,Husimi} for details. In this work, we will confine ourselves to the sudden switch and the adiabatic driving  protocols for which 
$\lambda=(\omega_c^2+\omega_h^2)/2\omega_c\omega_h$ and $\lambda=1$, respectively. 
The expressions for  mean heat exchanged during the hot and cold isochores can be evaluated, respectively, as follows
 \begin{eqnarray}
   Q_h   &=& \langle H \rangle_C-\langle H \rangle_B \nonumber
  \\
  &=& \frac{ \omega_h}{2}\Big[\text{coth}\Big(\frac{\beta_h \omega_h}{2}\Big) - \lambda_{AB} \text{coth}\Big(\frac{\beta_c  \omega_c}{2}\Big) \Big],  \label{heat2}
 \end{eqnarray}
  \begin{eqnarray}
  Q_c   &=& \langle H \rangle_A-\langle H \rangle_D \nonumber
  \\
  &=& \frac{  \omega_c}{2}\Big[\text{coth}\Big(\frac{\beta_c \omega_c}{2}\Big) - \lambda_{CD} \text{coth}\Big(\frac{\beta_h\omega_h}{2}\Big) \Big]. \label{heat4}   
 \end{eqnarray}
We are employing a sign convention in which all the incoming fluxes (heat and work) are taken to be positive.

\section{Asymmetric Quantum Otto heat engine}\label{sec3}

Let us first focus on the case in which the quantum Otto cycle (having as a working medium a time-dependent modulated harmonic oscillator) is working as a heat engine.
Since the working medium returns to its initial state after one complete cycle, the net work done on the system in a cycle is given by the first law of thermodynamics, $W=-(Q_h+Q_c)$. Work is said to be extracted from the engine when $W_{\rm ext}=-W=Q_h+Q_c>0$.  Accordingly, the efficiency of the engine is given by \cite{VOzgur2020}
\begin{equation}
\eta\!=\frac{W_{\rm ext}}{Q_h}   
\!=\!
1-\frac{\omega_c}{\omega_h}\frac{\text{coth}(\beta_c   \omega_c/2)- \lambda_{AB} \text{coth}(\beta_h\omega_h /2)}{\lambda_{CD} \text{coth}(\beta_c  \omega_c/2)-\text{coth}(\beta_h  \omega_h /2)} \, , \label{eq:efficiency-general}
\end{equation}
such that, when both the work strokes are adiabatic, Eq. \eqref{eq:efficiency-general} reduces to the simple form
\begin{equation}
\eta_\mathrm{ad}\!= \,  1 - \frac{\omega_c}{\omega_h}. \label{eq:efficiency-fully adiabatic}
\end{equation}
Our goal will be to study the efficiency of the engine in the cases in which one of the two work strokes \textit{is not} adiabatic. 
 
\subsection{Sudden Expansion Stroke}
Let us start by considering the case in which the expansion stroke, $C\rightarrow D$, is taken to be a sudden switch, thus involving friction in the operation of the heat engine. The compression stroke $A\rightarrow B$, instead, is still assumed to be adiabatic. The expression for extracted work $W_{\rm ext}$ (From now on, we will drop the subscript ``ext" and replace it with either SE or SC to specify sudden expansion and sudden compression strokes, respectively. In the following, all the expressions for work stand for extracted work.) and efficiency $\eta$ of the engine can be obtained by using Eqs. (\ref{heat2}) and (\ref{heat4}):
\bw
\be
W_{\rm SE}= \frac{(\omega_h-\omega_c)\left[(\omega_c+\omega_h) \coth(\beta_h\omega_h/2) 
-
					2\omega_h   \coth(\beta_c\omega_c/2) \right]         }
					{4\omega_h} \label{WCD},
\ee
and
\be
\eta_{\rm SE}= \frac{(\omega_h-\omega_c)\left[(\omega_c+\omega_h) \coth(\beta_h\omega_h/2) 
-
					2\omega_h  \coth(\beta_c\omega_c/2) \right]         }
					{2\omega_h^2  \left[\coth(\beta_h\omega_h/2)  - \coth(\beta_c\omega_c/2)  \right]}. \label{etaCD}
\ee
\ew
In deriving the expression for $W_{\rm SE}$ in Eq. (\ref{WCD}), we have substituted $\lambda_{AB}=1$ and $\lambda_{CD}=(\omega_c^2+\omega_h^2)/2\omega_c\omega_h$ into Eqs. (\ref{heat2}) and (\ref{heat4}), respectively, and then substituted the resulting expressions for $Q_h$ and $Q_c$ into $W_{\rm ext}=Q_h+Q_c$. The expression for the efficiency in Eq. (\ref{etaCD}), $\eta_{\rm SE}=1+Q_c/Q_h$, is obtained similarly. 

Eq. (\ref{etaCD}) can be recast into the following form:
\be
\eta_{\rm SE}=  \frac{1}{2}\left(1-\frac{\omega_c^2}{\omega_h^2}\right)  
\Bigg[
\frac
{
  \frac{\coth(\beta_h\omega_h/2)}{\coth(\beta_c\omega_c/2) } - \frac{2\omega_h}{\omega_c+\omega_h}  }
 {
  \frac{\coth(\beta_h\omega_h/2)}{\coth(\beta_c\omega_c/2) } - 1  } \Bigg].  \label{etaCD2}
\ee
The Positive Work Condition (PWC) $W_{\rm SE}\geq 0$ -- \textit{i.e.} the condition that the cycle is effectively behaving as an engine -- further requires
\be
  \frac{\coth(\beta_h\omega_h/2)}{\coth(\beta_c\omega_c/2) } \geq \frac{2\omega_h}{\omega_c+\omega_h} \geq 1, \label{PWC}
\ee
where the last inequality makes use of the assumption that $\omega_c \leq\omega_h$. Using Eq. (\ref{PWC}), one can prove that the term inside the square bracket in Eq. (\ref{etaCD2}) is always smaller than 1. Further, the condition $0 \leq 1-\omega_c^2/\omega_h^2\leq 1$ provides the non-trivial bound:
\be
\label{eq:sudden_expansion_loose_bound}
\eta_{\rm SE} \leq \frac{1}{2}.
\ee

Interestingly, we see that this bound is \textit{independent} of the frequencies, $\omega_c$ and $\omega_h$. This observation, together with Eq.~\eqref{eq:efficiency-fully adiabatic}, implies that the detrimental effect of the sudden switch is particularly strong in the limit $\omega_c \ll \omega_h$. In fact,  when both the work strokes are adiabatic and  $\omega_c \ll \omega_h$, the efficiency of the engine is close to $1$ and so we see that the effect of the sudden quench in the expansion stroke halves the efficiency, at least.

As is explained below, this is due to the friction introduced by the expansion stroke. In the sudden expansion stroke, the sudden quench of the frequency of the oscillator induces unwanted transitions between its energy levels and the system ends up in a non-equilibrium state. In the eigenbasis of the instantaneous Hamiltonian, the off-diagonal elements of the density matrix, dubbed as coherences, are non-zero.  Generating coherences via nonadiabatic driving costs an extra amount of energy (work) as compared to the adiabatic driving case, and as a result,  the system ends up storing additional parasitic energy. During the isochoric process that follows the sudden expansion, this additional energy cost is dissipated to the reservoirs. This phenomenon is referred to as inner or quantum friction \cite{Rezek2010,Plastina2014,Rezek2017,Feldmann2000,Feldmann2006,Ozgur2017}, and it negatively affects the engine's performance.

We can also immediately show that the Otto cycle with sudden expansion cannot work as a heat engine when working in the low-temperature limit. To see this, we notice that in the low-temperature limit, \textit{i.e.} when the conditions $\beta_k\omega_k \gg 1$,  $k=c, h$ are satisfied, we can approximate  $\coth(\beta_k\omega_k)=1$, such that the PWC in Eq. (\ref{PWC}) reduces to:
\begin{equation}
\label{eq:no_engine_low_temp_one}
   % \omega_c + \omega_h \geq 2\omega_h, 
   1 \geq \frac{2 \omega_h}{\omega_c + \omega_h} \, ,
\end{equation}
 which is incompatible with the assumption $\omega_c \leq \omega_h$, except for the trivial case $\omega_c=\omega_h$. There is a physical explanation for this: in the low-temperature regime, for a typical thermal state only the lowest levels of the harmonic oscillator are populated. When we drive the system non-adiabatically by using the sudden switch protocol, most of the transitions will be transitions from lower energy states to higher energy states, which will effectively raise the temperature of the harmonic oscillator as compared to the adiabatic case and, as a result, a larger amount of heat will be dissipated to the cold reservoir. If the amount of heat dissipated to the cold reservoir exceeds the amount of heat extracted from the hot reservoir, the Otto cycle cannot work as a heat engine. By extending the same logic to all temperature regimes, we can conclude that to minimize the frictional effects, we should operate the engine under consideration in the high-temperature regime. In the high-temperature regime, the upper levels of the harmonic oscillator will also be appreciably populated. Thus, the non-adiabatic transitions will be in both directions -- upwards and downwards -- resulting in an effectively lower temperature of the oscillator as compared to the previous low-temperature regime discussion.

 % Hence, in the low-temperature regime, the Otto cycle with sudden expansion stroke  cannot function as a heat engine given the highly frictional nature of the sudden switch protocol. The same is true for the harmonic Otto cycle with sudden compression stroke and we will show this in Sec. IV.

%
\subsection{Upper bound on the efficiency}
In the previous subsection, we found that the efficiency of the engine is bounded from above by $1/2$. However, we do not know whether this bound is tight or loose. In the following, we find an expression for the maximum efficiency of the engine in the high-temperature regime. We also conjecture (supported by heuristic as well as numerical checks) that this maximum value can be used as a bound -- tighter than Eq. \eqref{eq:sudden_expansion_loose_bound} -- for the maximal efficiency of the engine in any regime.

\begin{figure}
 \centering
\includegraphics[width=8.6cm]{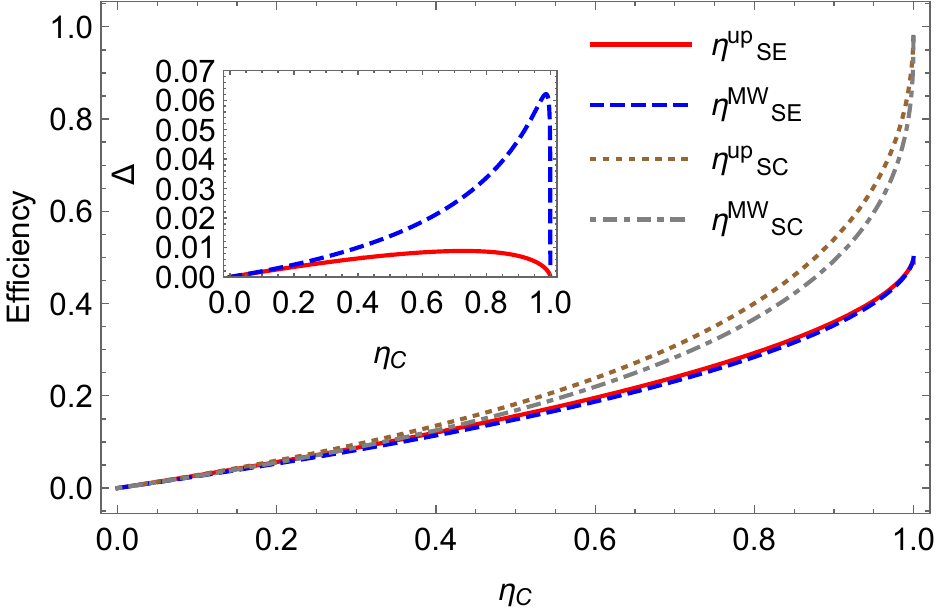}  
 \caption{Plots of $\eta^{\rm up}_{\rm SE}$ (Eq. (\ref{AuxB})),  $\eta^{\rm MW}_{\rm SE}$ (Eq. (\ref{AuxD})),  $\eta^{\rm up}_{\rm SC}$ (Eq. (\ref{AuxC})) and  $\eta^{\rm MW}_{\rm SC}$ (Eq. (\ref{EMW2})) versus Carnot efficiency $\eta_C$. In the inset, we have plotted the differences, $\Delta=\eta^{\rm up}_{\rm SE}-\eta^{\rm MW}_{\rm SE}$ (solid red curve in the inset) and $\Delta'=\eta^{\rm up}_{\rm SC}-\eta^{\rm MW}_{\rm SC}$ (dashed blue curve in the inset).}
\label{fig:eta_up}
\end{figure}

In the high-temperature limit, which is often employed to get analytic results while studying quantum heat engines \cite{VJ2019,VJ2020,Geva1992,Geva1994,MePRR,VRosa2023}, \textit{i.e.} in the limit $\beta_k\omega_k/2 \ll 1$, we can set $\coth(\beta_k\omega_k/2)\approx 2/\beta_k \omega_k\,(k=c,\,h)$. Then the expressions for the extracted work and efficiency take the following forms
\bea
W^{\rm HT}_{\rm SE} &=&  (1-z) \left(\frac{z+1}{2}-  \frac{\tau }{z}\right)\frac{1}{\beta_h} , \label{WCDHT}
\\
\eta^{\rm HT}_{\rm SE} &=& \frac{(1-z) \left(z^2+z-2\tau\right)}{2 (z-\tau )}, \label{etaCDHT}
\eea
where we defined $z\equiv\omega_c/\omega_h$ and $\tau \equiv \beta_h/\beta_c$. The PWC implies that
\be
W^{\rm HT}_{\rm SE}  \geq 0 \Rightarrow \frac{z^2+z}{2} \geq \tau,    \label{PWCSE}
\ee
which is a more stringent condition for extracting work from the engine cycle as compared to the engine cycle in which both work strokes are adiabatic in nature, and the PWC is simply given by $z>\tau$. Since $(z^2+1)/2<z$,  for the given  $\tau$ (ratio of the temperatures of the cold and hot reservoirs), the compression ratio $z$ for the sudden expansion stroke should be significant when compared to its adiabatic counterpart in order to extract positive work from the engine cycle.

\begin{comment}
Eq. \eqref{etaCDHT} can be maximized as a function of the compression rate $z$. By replacing  $z$ with the value for $z$ which maximizes \eqref{etaCDHT}, denoted $z^\ast$, we get an expression for the maximum efficiency of the engine in terms of the temperature ratio, $\tau$. The explicit expression for the optimal efficiency, which we denote as $\eta_{\mathrm{SE}}^\mathrm{up}$, is not particularly illuminating, and we limit ourselves to reporting its behavior as a function of the Carnot efficiency, $\eta_C \equiv 1 - \tau $, in Fig. \ref{fig:eta_up} (solid red curve).
\end{comment}
Eq. \eqref{etaCDHT} can be maximized as a function of the compression rate $z$. By setting $\partial \eta^{\rm HT}_{\rm SE}/\partial z=0$, we have 
\begin{equation}
     2z^3 -3 z^2\tau +\tau (2\tau-1)=0 . \label{CI1}
\end{equation}
The above equation, being a case of \textit{casus irreducibilis} (see Appendix A), cannot be solved in real radicals. However, a solution in terms of trignometric functions is available, and is given by
\begin{equation}
 z^* =  \frac{\tau }{2} +  \tau \cos \left(\frac{1}{3} \cos ^{-1}\left(\frac{(\tau -4) \tau +2}{\tau^2 }\right)\right). \label{auxA}
\end{equation}
Substituting Eq. (\ref{auxA}) in Eq. (\ref{etaCDHT}), the expression for the maximum efficiency is found to be
\begin{equation}
 \eta^{\rm up}_{\rm SE} =  \frac{(2 K+\tau -2) \big[4 K^2+4 K (1+\tau)-(6-\tau) \tau \big]}{8\tau -16 K} \label{AuxB},
\end{equation}
where $K = \tau \cos \left(\frac{1}{3} \cos ^{-1}\left(\frac{(\tau -4) \tau +2}{\tau^2 }\right)\right)$. Eq. (\ref{AuxB}), which is plotted in Fig. 2 (see solid red curve),  is not particularly illuminating. 
Instead -- to gain some insights into how the friction affects the efficiency -- we provide the expansion of $\eta^{\rm up}_{\rm SE}$  as Taylor's series in $\eta_C$.
The expansion up to the cubic order reads, 
\be
 \eta^{\rm up}_{\rm SE} = (2-\sqrt{3})\eta_C + \left(\sqrt{3}-\frac{5}{3}\right)\eta_C^2 + \frac{\eta_C^3}{18\sqrt{3}} + \dots . \label{Taylor}
\ee
We will comment more on this expansion in the next section.
Although the upper bound on the efficiency, $\eta^{\rm up}_{\rm SE}$, has been obtained in the high-temperature limit, one can argue that its validity is more general and that it can be promoted to be an upper bound of the efficiency valid in any temperature regime. Heuristically, this statement can be motivated by observing, as we did in Eq.\eqref{eq:no_engine_low_temp_one}, that in the low-temperature limit, the Otto cycle cannot be used as an engine tout-court. Hence, it is reasonable to expect that the efficiency of the engine \textit{reduces} when reducing the temperature and so the bound derived in the high-temperature limit can be extended to finite temperature as well.

 To further substantiate this argument, we plot a histogram, for given finite temperatures, of the sampled values of efficiency, given in Eq. (\ref{etaCD}) with random samplings of the parameters ($\omega_c, \omega_h$). 
 
\begin{figure}
 \centering
\includegraphics[width=8.6cm]{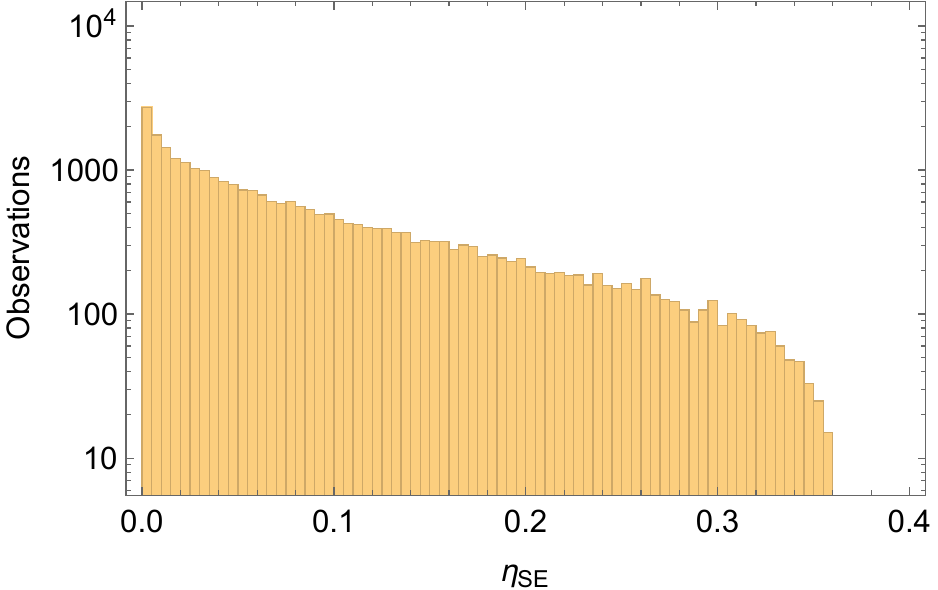}  
 \caption{Histogram of sampled values of $\eta_{\rm SE}$ given in Eq. (\ref{etaCD}) for random sampling over a region of the parametric space 
 ($\omega_c,\,\omega_h$). The parameters are sampled over the uniform distributions $\omega_{c, h}\in [0,100]$ at fixed values of $\beta_c=1$ and $\beta_h=1/10$.  For plotting the histograms, we choose a bin width of 0.01 to arrange $10^6$ data points.}
 \label{fig:histogram_expansion}
\end{figure}

 For the chosen values of $\beta_c=1$ and $\beta_h=1/10$, we have $\eta^{\rm up}_{\rm SE}=0.36$. As we can see in Fig. \ref{fig:histogram_expansion}, all the sampled values of efficiency $\eta_{\rm SE}$ lie below this bound, which supports our conjecture that $\eta^{\rm up}_{\rm SE}$ serves as an upper bound in all operational regimes. While uniformly sampling the parametric space ($\omega_c,\,\omega_h$), we
choose $\omega_{1, 2}\in [0,100]$ so that the parametric space spans all operational regimes, including regimes far away from being in the high-temperature regime.

\subsection{Efficiency at maximum work}
We can also obtain the expression for the efficiency at maximum work output by optimizing Eq. (\ref{WCDHT}) with respect to $z$. The resulting expression for the efficiency is given by
\bea 
 \eta^{\rm MW}_{\rm SE}  &=& \frac{3 \sqrt[3]{1-\eta_C}-2 \eta_C \sqrt[3]{1-\eta_C}+3 \eta_C-3}{2 \left(\sqrt[3]{1-\eta_C}+\eta_C-1\right)} \label{AuxD}
 \\
 &=&  \frac{\eta_C}{4} +\frac{5 \eta_C^2}{72}    + \frac{5 \eta_C^3}{144} + \dots \label{EMW1}
 \eea
 We note that the linear order term in $\eta_C$ in Eq. (\ref{EMW1}) is given by $\eta_C/4$, while for the adiabatic harmonic Otto engine,  the linear order term in $\eta_C$ is given by $\eta_C/2$, which in turn indicates that adiabatic engine is far more efficient than the engine with sudden expansion stroke. In Fig. \ref{fig:eta_up}, we have plotted $ \eta^{\rm MW}_{\rm SE}$ (dashed blue curve) as a function of $\eta_C$. In the inset, we have plotted the difference between $ \eta^{\rm up}_{\rm SE}$ and  $\eta^{\rm MW}_{\rm SE}$. It is clear that the difference $\Delta=\eta^{\rm up}_{\rm SE}-\eta^{\rm MW}_{\rm SE}$ is very small.  This indicates that the engine's performance is dominated by frictional effects, thus rendering maximum efficiency and efficiency at maximum work values very close to each other.

\section{Harmonic Otto engine with sudden compression stroke}\label{sec4}

Having studied the sudden expansion stroke case, we now turn our attention to the modified scenario in which the compression stroke is instantaneous. The expression for the work output can be derived by substituting $\lambda_{AB}=(\omega_c^2+\omega_h^2)/2\omega_c\omega_h$ and  $\lambda_{CD}=1$ in Eqs. (\ref{heat2}) and (\ref{heat4}), respectively, and then substituting the resulting expressions for $Q_h$ and $Q_c$ in $W_{\rm ext}=Q_h+Q_c$. Similarly, the expression for the efficiency, $\eta_{\rm SE}=1+Q_c/Q_h$, can be obtained. The final expressions read
\bw
\bea
W_{\rm SC} &=& \frac{\left(\omega_c-\omega_h\right) \left[\left(\omega_c+\omega_h\right) \coth \left(\frac{\beta _1 \omega_c}{2}\right)-2 \omega_c \coth \left(\frac{\beta_h \omega_h}{2}\right)\right]}{4 \omega_c}, \label{WSC}
\\
\eta_{\rm SC} &=& \frac{\left(\omega_c-\omega_h\right) \left(\left(\omega_c+\omega_h\right) \coth \left(\frac{\beta _1 \omega_c}{2}\right)-2 \omega_c \coth \left(\frac{\beta_h \omega_h}{2}\right)\right)}
{2 \omega_c \omega_h \coth \left(\frac{\beta_h \omega_h}{2}\right)-\left(\omega_c^2+\omega_h^2\right) \coth \left(\frac{\beta _1 \omega_c}{2}\right)}. \label{etaSC}
\eea
 \ew
Contrary to the previous case, Eqs. (\ref{WSC}) and  (\ref{etaSC}) \textit{do not} provide any non-trivial bound on the efficiency, $\eta_{\rm SC}$. Indeed, by direct inspection, one can see that Eqs. (\ref{WSC}) and  (\ref{etaSC}) simply lead to the trivial bound $\eta_{\rm SC}\leq 1$.
However, frictional effects still play a role here in reducing the efficiency of the engine, as we will show momentarily.  

Similarly to the sudden expansion case, it is possible to show that the Otto cycle does not work as an engine in the low-temperature limit. Indeed, the PWC from Eq. (\ref{WSC}) reads 
\be
(\omega_c+\omega_h) \coth \left(\frac{\beta _1 \omega_c}{2}\right)\leq 2\omega_c \coth\left(\frac{\beta_h\omega_h}{2}\right), \label{AA1}
\ee
which  in the low-temperature regime reduces to the condition $\omega_c+\omega_h\leq 2\omega_c$, which is incompatible with the assumption $\omega_h \geq \omega_c$, except for the trivial case $\omega_h=\omega_c$.

\subsection{Upper bound on the efficiency with sudden compression stroke}

Eqs. (\ref{WSC}) and  (\ref{etaSC}) simplify dramatically when the high-temperature limit is considered.
Explicitly, they reduce to
\bea
W^{\rm HT}_{\rm SC} &=& (z-1) \left(\frac{\tau  (z+1)}{2 z^2}-1\right)\frac{1}{\beta_h}, \label{WSCHT}
\\
\eta^{\rm HT}_{\rm SC} &=&  \frac{(z-1) \left(2 z^2-\tau  (z+1)\right)}{\tau +(\tau -2) z^2}, \label{etaSCHT}
\eea
where $z=\omega_c/\omega_h$ and $\tau=\beta_h/\beta_c$. 

Once again, the upper bound on the efficiency can be found by maximizing Eq. (\ref{etaSCHT}) with respect to the compression ratio $z$, as a function of $\tau$. The resulting expression for $ \eta^{\rm up}_{\rm SC}$, in terms of trigonometric functions, is given by
\begin{equation}
  \eta^{\rm up}_{\rm SC} =\frac{ 2-\tau + 16 \sqrt{\tau/(2-\tau)} \cos ^3 B-4 (2+\tau) \cos ^2 B}{(\tau -2) [2 \cos (2 B)+1]}, \label{AuxC}
\end{equation}
where $B=\frac{1}{3} \cos ^{-1}\left(-\sqrt{(2-\tau ) \tau }\right)$.
 Again, as s we did for the sudden expansion case, we present the  Taylor expansion of the above equation up to the third order term in $\eta_C$:
\begin{figure}
 \centering
\includegraphics[width=8.6cm]{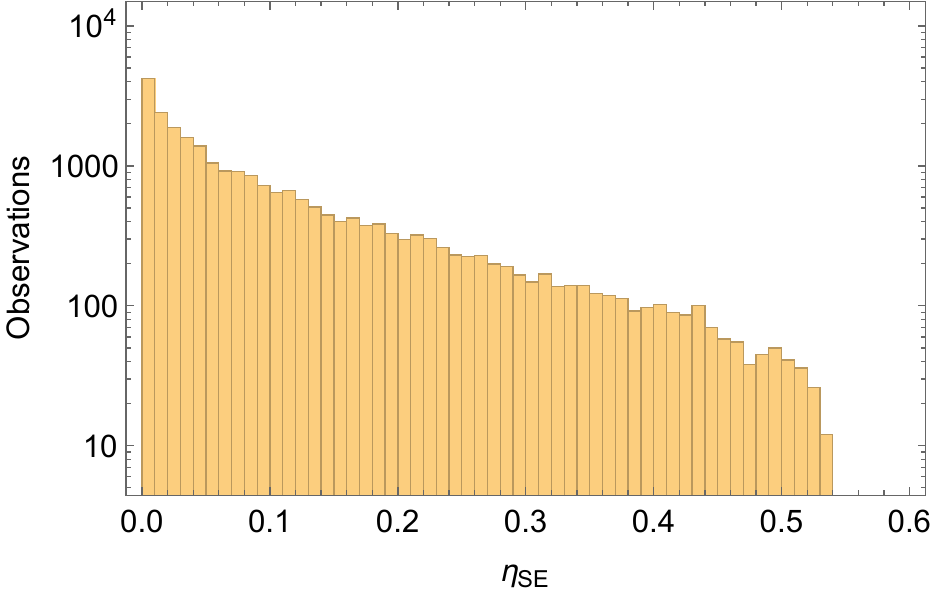}  
 \caption{Histogram of sampled values of $\eta_{\rm SC}$ given in Eq. (\ref{etaSC}) for random sampling over a region of the parametric space 
 ($\omega_c,\,\omega_h$). The parameters are sampled from uniform distributions $\omega_{c,h}\in [0,100]$ at fixed values of $\beta_c=1$ and $\beta_h=1/10$. The histograms have a bin width of 0.01 to arrange $10^6$ data points.}
\end{figure}
\bw
\be
 \eta^{\rm up}_{\rm SC}  = (2-\sqrt{3})  \eta_C +  \frac{2}{3} \left(3\sqrt{3}-5\right)\eta_C^2 + \frac{1}{54} \left(252-143 \sqrt{3}\right) \eta_C^3 + \dots \label{Taylor2}
\ee
\ew

Eq. (\ref{Taylor2}) must be compared with its analogous result in the case of the sudden expansion stroke,   Eq. (\ref{Taylor}). 
Interestingly, we see that the two expressions are \textit{identical} on their linear terms. However, the symmetry between the two cases is broken by the second-order term in the $\eta_C$ expansion. In this case, we see that the coefficient for the sudden compression, Eq. \eqref{Taylor2}, is \textit{exactly twice} its counterpart in Eq. \eqref{Taylor}.

All-in-all, the analysis of the Taylor expansion in $\eta_C$ shows two remarkable and related phenomena: the sudden expansion and the sudden compression strokes \textit{are not} equivalent and, when working in the high-temperature limit in particular, the sudden compression protocol outperforms the sudden expansion protocol. Moreover, the asymmetry between the two configurations arises from the second-order term in the expansion in $\eta_C$, with the sudden compression stroke coefficients being exactly twice the same coefficient appearing in the expansion for the sudden expansion case. 

To conclude the analysis of Eq. (\ref{AuxC}) , we plot its behavior in Fig. 2 (dotted brown curve). It is clear from the figure that $\eta^{\rm up}_{\rm SC}>  \eta^{\rm up}_{\rm SE}$, and unlike $ \eta^{\rm up}_{\rm SE}$ (solid red curve in Fig. 2) which can attain $1/2$ as its maximum value, $ \eta^{\rm up}_{\rm SC}$ attains unit efficiency for $\eta_C=1$.
Once again, we claim that the non-trivial bound of Eq. (\ref{AuxC}), although obtained in the high-temperature limit, extends its validity to the finite temperature case. To support this claim, we recall that the Otto cycle does not work as an engine in the low-temperature limit as proven earlier. Moreover, exactly as we did for the sudden expansion case,
we have plotted a histogram (see Fig. 4) for the sampled values of $\eta_{\rm SC}$ (see Eq. (\ref{etaSC})) for random sampling over a region of parametric space ($\omega_c,\omega_h$). Here, for the  chosen values $\beta_c=1$ and $\beta_h=10$, the upper bound on the efficiency is given by $\eta^{\rm up}_{\rm SE}=0.54$. Clearly,  for the given values of $\beta_c$ and $\beta_h$,  all sampled values of $\eta_{\rm SC}$  lie below  $\eta^{\rm up}_{\rm SC}$, in agreement with the claim that $\eta^{\rm up}_{\rm SC}$ is an upper bound at all temperatures.

\subsection{Efficiency at maximum work}
Let us now evaluate the analytic expression of efficiency at maximum work. Optimization of Eq. (\ref{etaSCHT}) with respect to $z$ yields the following expression
\bw
\be
\eta^{\rm MW}_{\rm SC} = -\frac{4 \left(\sqrt[3]{1-\eta _C}+\left(1-\eta _C\right){}^{2/3}-2\right)+\eta _C \left(4 \sqrt[3]{1-\eta _C}+\eta _C-1\right)}{4 + \eta _C \left(\eta _C+3\right)}
=
  \frac{\eta _C}{4}     +   \frac{17 \eta _C^2}{144}  +    \frac{41 \eta _C^3}{576} + \dots  \label{EMW2}
\ee
\ew
Again, we notice that $\eta^{\rm MW}_{\rm SC}$  and $\eta^{\rm MW}_{\rm SE}$ (see Eq. (\ref{EMW1})) share the first term  $\eta_C/4$ in their respective Taylor series. The differences start to appear from the second order term in $\eta_C$. We plot Eq. (\ref{EMW2}) in Fig. 2 (see dot-dashed gray curve). It just lies below the curve (dotted brown curve) for $\eta^{\rm up}_{\rm SC}$. In the inset of Fig. 2, we have plotted the difference between 
$\eta^{\rm up}_{\rm SC}$ and $\eta^{\rm MW}_{\rm SC}$, $\Delta'=\eta^{\rm up}_{\rm SC}-\eta^{\rm MW}_{\rm SC}$ (dashed blue curve in the inset of Fig. 2).
\begin{figure}
 \centering
\includegraphics[width=8.6cm]{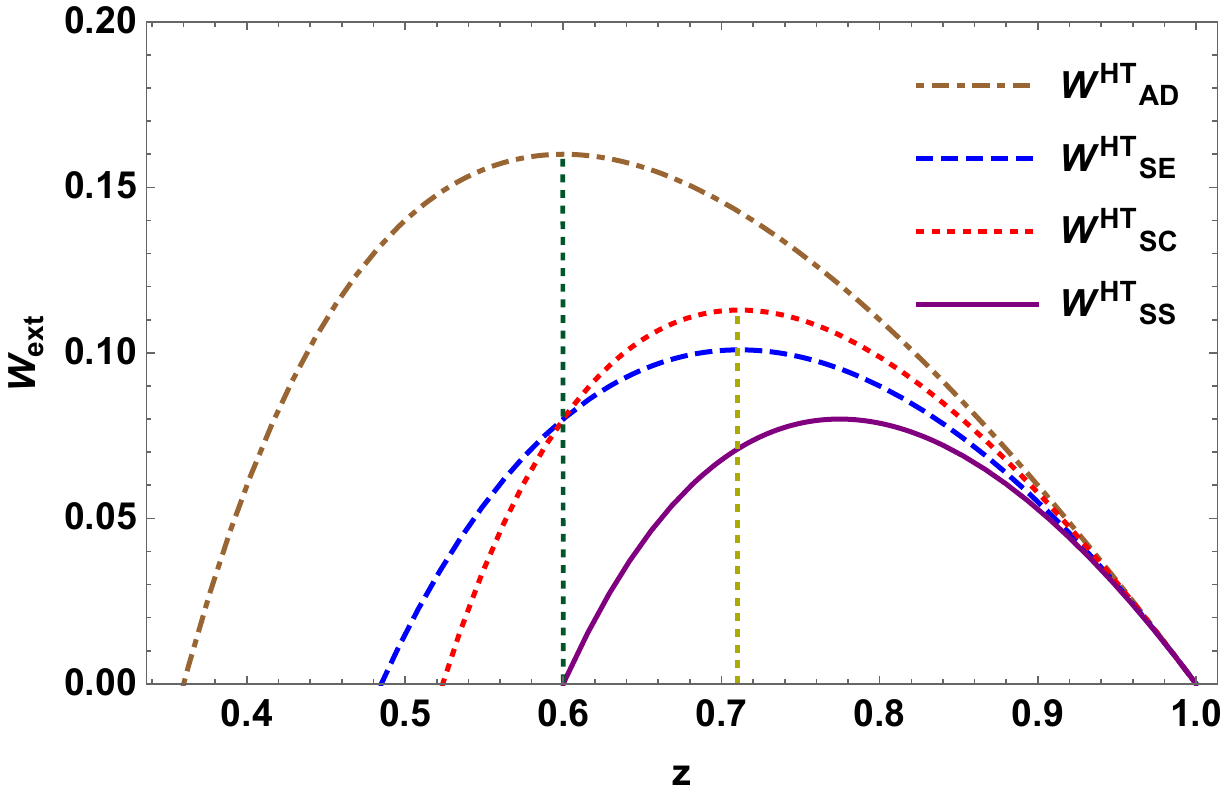}  
 \caption{Work output as a function of compression ratio $z$ for various cases discussed in the main text. Dot-dashed brown, dashed blue, dotted red and solid purple curves represent the adiabatic ($W^{\rm HT}_{\rm AD}=(1-z)(1-\tau/z)$), sudden expansion (Eq. (\ref{WCDHT})), sudden compression  (Eq. (\ref{WSCHT})), and sudden-switch case ($W^{\rm HT}_{\rm SS}=(1-z^2) (z^2-\tau)/2 z^2=0$), respectively. Here, we have fixed $\tau=0.36$. The intersection between dashed blue and dotted red curves is given by $z=\sqrt{\tau}=\sqrt{0.36}=0.6$. It is also clear from the plots that both the sudden compression case and the sudden expansion case exhibit a maximum of work at the same value of $z$ given by $z=\tau^{1/3}=0.711$.} \label{workoutput}
\end{figure}

\section{Comparison}\label{sec5}
Having presented the results for two different examples of asymmetric Otto cycles, we now perform a more in-depth comparison of their performances. Moreover, we point out some interesting connections to the case of the \emph{symmetric} Otto heat engine, considering the cases in which both strokes are sudden and/or both strokes are adiabatic. 

In Fig. \ref{workoutput} we plot the work output for all the four cases mentioned above -- sudden expansion, sudden compression, symmetric sudden (aka sudden switch), symmetric adiabatic -- against the compression ratio $z$ of the harmonic oscillator, with a fixed value of $\tau=0.36$ \footnote{We performed extensive numerical checks that the results are not qualitatively affected by changing the value of $\tau$}. 
We note that the curves for sudden expansion and sudden compression strokes intersect when the condition $z=\sqrt{\tau}$ is met. This condition can be derived analytically, by equating $W^{\rm HT}_{\rm SE}$ and $W^{\rm HT}_{\rm SC}$, and in Fig. \ref{workoutput} it corresponds to the point $z = 0.6$. 
Interestingly, we notice that this condition, $z=\sqrt{\tau}$, precisely corresponds to the condition of no work output for the case in which both strokes are sudden, $W^{\rm HT}_{\rm SS}=(1-z^2) (z^2-\tau)/2 z^2=0$ \cite{VOzgur2020}.
% This is interesting as for harmonic  Otto engine in which both adiabatic branches are sudden,  $z=\sqrt{\tau}$ represent the point where the work output $W^{\rm HT}_{\rm SS}=(1-z^2) (z^2-\tau)/2 z^2=0$ \cite{VOzgur2020}. The positive work condition in sudden switch case implies that $z$ should be greater than $\sqrt{\tau}$ for extracting from from the Otto cycle. 
In addition, we notice that the point $z=\sqrt{\tau}$ also corresponds to the value of $z$ at which work extraction ($W^{\rm HT}_{\rm AD}=(1-z)(1-\tau/z)$) is maximum when the Otto cycle is adiabatically driven.

Another interesting observation is that when the Otto cycle contains only one of a sudden expansion or a sudden compression stroke, the work extraction is maximum for $z=\tau^{1/3}$ (see light green vertical line in Fig. \ref{workoutput} at $z=\tau^{1/3}=0.36^{1/3}=0.711$).

\begin{figure}[H]
 \centering
\includegraphics[width=8.6cm]{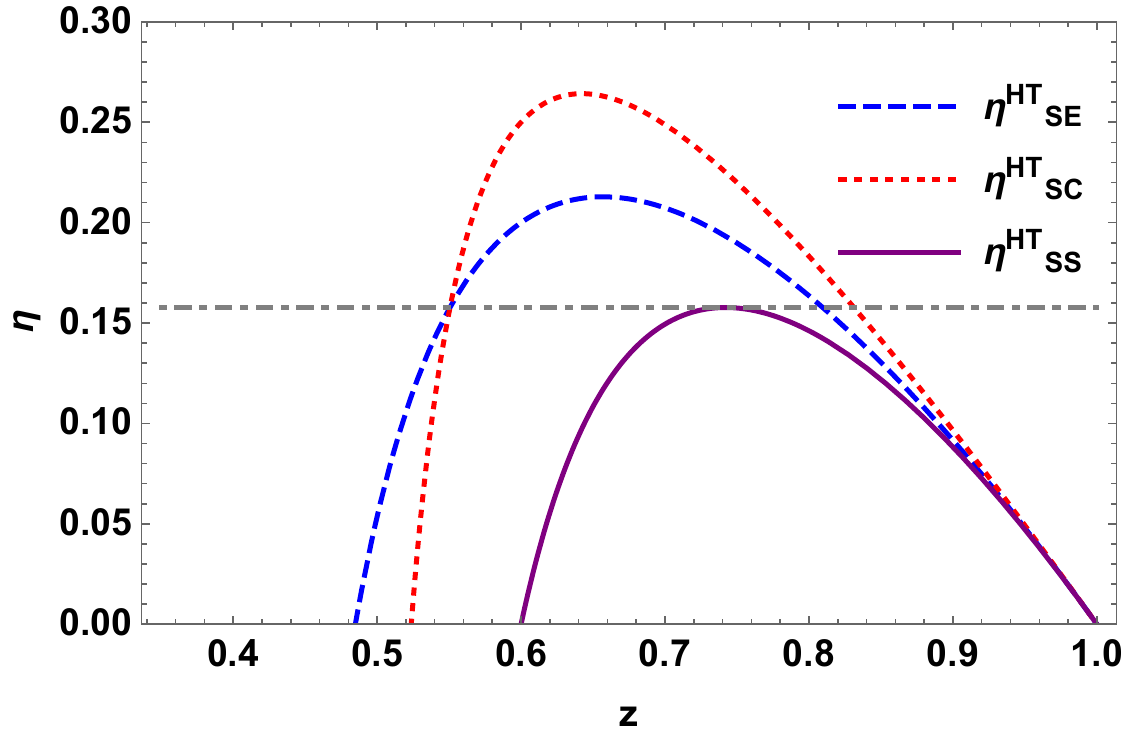}  
 \caption{Efficiency as a function of compression ratio $z$ for three different cases involving at least one sudden stroke as discussed in the main text. Dashed blue, dotted red, and solid purple curves represent the sudden expansion (Eq. (\ref{etaCDHT})), sudden compression  (Eq. (\ref{etaSCHT})), and sudden-switch case ($\eta^{\rm HT}_{\rm SS}=(z^2-1) (z^2-\tau)/(\tau- z^2(2-\tau))$), respectively. Here, we have fixed $\tau=0.36$. At the intersection point between dashed blue and dotted red curves, the efficiency is given by  $\eta_{\rm intsec}=(3-2\sqrt{2(1-\eta_C)}-\eta_C)\eta_C/(1+\eta_C)^2=0.158$, which corresponds to the maximum efficiency achievable in the sudden-switch case. } \label{effsz}
\end{figure}

For completeness, we also plot the efficiency as a function of the compression ratio $z$ in Fig. \ref{effsz} for sudden expansion, sudden compression, and sudden switch cases. We observe that the value of the efficiency at which the curves for sudden expansion and sudden compression intersect, precisely agrees with the maximum efficiency (see horizontal dot-dashed line in Fig. \ref{effsz}) reached by the sudden switch case. The analytic expression for the maximum efficiency for the sudden switch case (as well as efficiency at intersection point in Fig. \ref{effsz}) was derived in Ref. \cite{VOzgur2020} and is given by $\eta^{\rm SS}_{\rm max} = \eta_{\rm intsec}=(3-2\sqrt{2(1-\eta_C)}-\eta_C)\eta_C/(1+\eta_C)^2$.

It is clear from Fig. \ref{workoutput} that for the range $(\sqrt{8\tau+1}-1)/2<z<\sqrt{\tau}$ (see also Eq. (\ref{PWCSE})), more work can be extracted for the case with sudden expansion stroke as compared to the case with sudden compression stroke, while the opposite is true for the regime $\sqrt{\tau}<z<1$. Similar conclusions about the efficiency can be drawn by looking at Fig. \ref{effsz}; with the only difference being that, in this case, the intersection point lies at $z=(\tau+\sqrt{2\tau}(1-\tau))/(2-\tau)$ \footnote{This solution can be obtained by putting  $\eta^{\rm HT}_{\rm SE}=\eta^{\rm HT}_{\rm SC}$ and solving for $z$.}.

\begin{figure}
 \centering
\includegraphics[width=8.6cm]{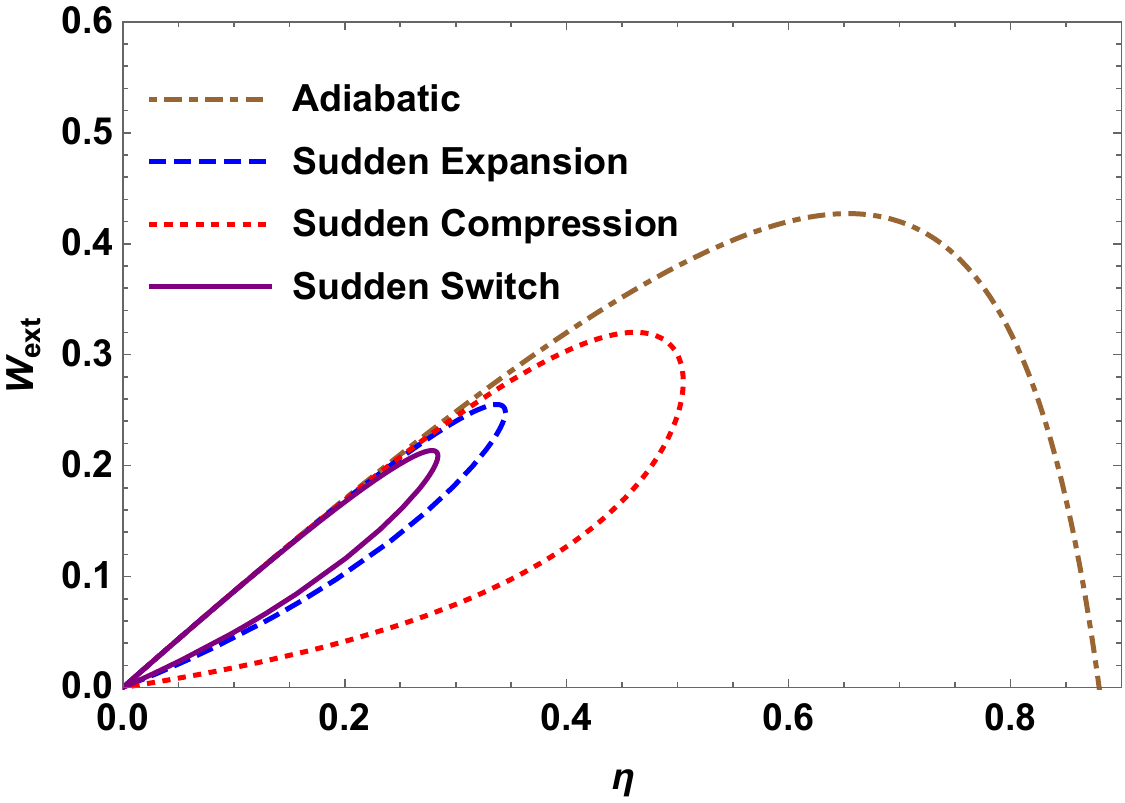}  
 \caption{Work-efficiency plots for all four cases that are discussed in the main text. Except for the adiabatic case, all curves are loop-like, characterizing the signature of the friction effect in the performance of the engine under consideration.} \label{loops}
\end{figure}

Finally, to find the optimal operational regime for each case, we plot efficiency-work curves in Fig. \ref{loops}. Except for the adiabatic case (which has an open-ended shape), the efficiency-work curves are loop-shaped curves for all cases involving nonadiabatic driving in at least one of the work strokes. The loop-like behavior characterizes the frictional effects arising due to the nonadiabatic nature of driving \cite{Gordon1991,Gordon1992,GordonKosloff,Benenti2017,VS2022}. In all three cases with nonadiabatic driving, the optimal operational regime is located on the portion of the curve having a negative slope. As is clear from Fig. \ref{loops}, a sudden switch driving in both adiabatic branches leads to an engine characterized by poor efficiency and less work extraction. 

If we compare the case of the sudden expansion stroke to that of the sudden compression stroke, it turns out that the performance of the latter is much better than the former.

The comparisons made in this Section allowed us to uncover a clear asymmetry between the overall efficiency of the Otto engine when just one of the two strokes is adiabatic, with the sudden compression engine outperforming the sudden expansion engine. It should be emphasized that such an asymmetry is not easy to predict based on heuristic arguments. At the same time, we also uncovered that, quite interestingly, the four cases have some unexpected relations between their performances for specific values of the compression ratio $z$, as discussed in Figs.~\ref{workoutput} and \ref{effsz}.

\section{Complete phase diagram of the quantum harmonic  Otto cycle}\label{sec6}

\begin{figure*}
 \centering
\includegraphics[width=17cm]{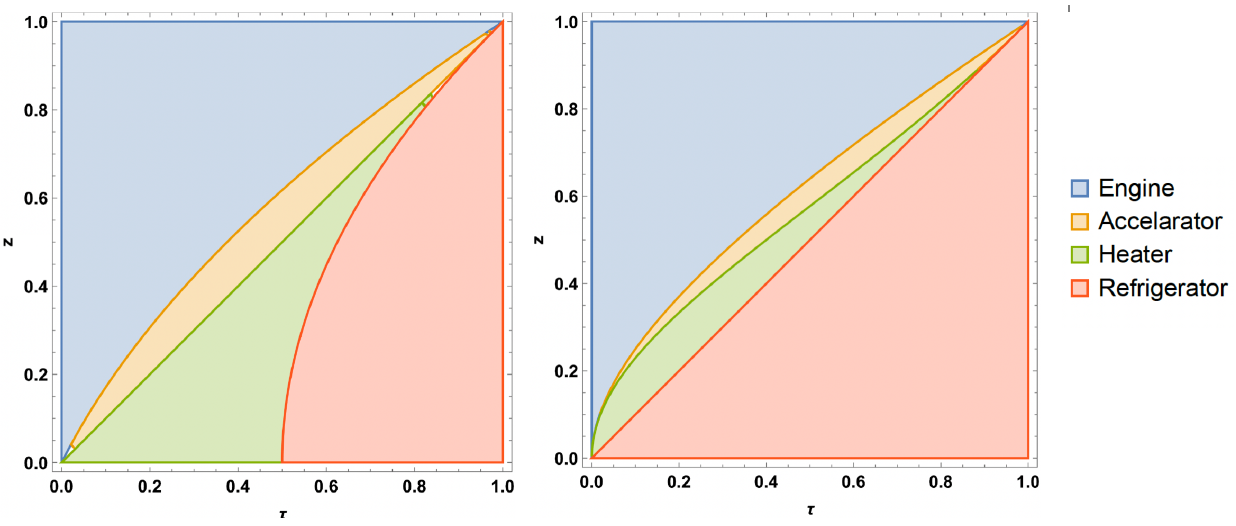}  
 \caption{Full phase diagram of the harmonic quantum Otto cycle as a function of compression ratio $z$ and ratio of cold to hot temperatures $\tau$. The phase diagram is obtained by imposing the conditions given in Eq. (\ref{conditions}) on  Eqs. (\ref{WCDHT}), (\ref{WSCHT}), (\ref{QhcSE})  and (\ref{QhcSC}). Left  (right) panel corresponds to the Otto cycle with sudden expansion  (compression) work stroke.   } \label{phase_diagram}
\end{figure*}

% \textcolor{red}{Frictional effects can alter the PWC, thereby modifying the parametric regime of the operation of Otto cycle as a heat engine or refrigerator CITATION NEEDED. }
As was already noted, the PWCs are affected when frictional effects are present \cite{VOzgur2020,Kiran2022}. This observation, in turn, motivates one to study the phase diagram of the Otto Cycle in the presence of frictional effects. As we will see, depending on the values of the parameters $z$ and $\tau$, the asymmetric Otto cycle can work not just as an engine or a refrigerator, but also in operational modes corresponding to a heater and thermal accelerator (the latter two being absent when frictional effects are not present).

In general, a thermal device can operate in the following four modes:
\begin{align}
 & \text{Engine}: & W_{\rm ext} \geq 0, Q_h \geq 0, Q_c \leq 0,\nonumber \\
 & \text{Refrigerator}: & W_{\rm ext} \leq 0, Q_h \leq 0, Q_c \geq 0, \nonumber \\
 & \text{Heater}: & W_{\rm ext} \leq 0, Q_h \leq 0, Q_c \leq 0, \nonumber \\ 
 & \text{Thermal~accelerator}: & W_{\rm ext} \leq 0, Q_h \geq 0, Q_c \leq 0,  \label{conditions}
\end{align}
In the engine mode, the system absorbs heat ($Q_h\geq 0$) from the hot reservoir, converts it into work ($W_{\rm ext}\geq 0$), and dumps the remaining heat ($Q_c\leq 0$) into the cold reservoir. In the refrigerator mode, work is utilized ($W_{\rm ext}\leq 0$) to transport heat from the cold ($Q_c\leq 0$) to the hot reservoir ($Q_h\leq 0$). In the heater mode, work is utilized  ($W_{\rm ext}\leq 0$) to dump heat at both the hot  ($Q_h\leq 0$) and the cold ($Q_c\leq 0$) reservoirs. In the thermal accelerator mode, work ($W_{\rm ext}\leq 0$)  is utilized to transport heat from the hot ($Q_h\geq 0$)  to the cold reservoir ($Q_c\leq 0$), which is otherwise a spontaneous process.

The heater and thermal accelerator modes, usually overlooked in the literature, can be potentially useful~\cite{Piccione2021PRA,Dieguez2023}. The purpose of a heater 
% (or heat pump) 
is to heat, simultaneously, both a hot and a cold reservoir.
% Therefore, the coefficient of performance (COP) of this thermal machine is the ratio of the magnitude of the heat given to the hot reservoir to the inputted work. In contrast, for a refrigerator, where the interest is to cool further the cold reservoir, the COP is the ratio of the heat taken up from the cold reservoir to the inputted work. 
Finally, for the thermal accelerator mode, a COP has been proposed as the ratio between heat delivered to the cold reservoir over the performed work~\cite{Piccione2021PRA,Dieguez2023}. One could envisage situations where it is useful to evacuate heat as fast as possible in order to cool hot spots, and in that 
context a quantum accelerator would be a machine that facilitates the natural heat flow from the hot bath to the cold bath. In that case, one should consider the heat extracted from the hot spot (hot reservoir) as useful output and the inputted work as a cost. 

In the following, we discuss the phase diagram for the two different versions of the asymmetric harmonic quantum Otto cycle: one with a sudden expansion stroke and the other with a sudden compression stroke. 
\\
\subsection{Asymmetric Otto cycle with sudden expansion stroke}
It is worth noting that when both of the work strokes are quasi-static in the Otto cycle under consideration, either we can achieve engine mode or refrigeration mode; the heater and accelerator modes are absent from the phase diagram of the quasi-static Otto cycle and the phase diagram is symmetric with respect to the parameter space available for the engine and refrigerator modes. However, when either or both of the work branches are driven in finite time, frictional effects arise and new operational regimes emerge in the phase diagram.

For the sudden expansion stroke, the expression for extracted work  is given in Eq. (\ref{WCDHT}).  In the high-temperature limit, the expressions for $Q_h$ and $Q_c$ can be obtained using Eqs. (\ref{heat2}) and (\ref{heat4}), and are given by
\begin{equation}
Q_{h(SE)}^{\rm HT} = \frac{1}{\beta_h}\left(1 - \frac{\tau}{z}\right), \quad Q_{c(SE)}^{\rm HT} =     \frac{1}{\beta_h}\left(\tau - \frac{1+z^2}{2}\right). \label{QhcSE}
\end{equation}
As can be seen from the PWC given in Eq. (\ref{PWCSE}), the asymmetric harmonic Quantum Otto cycle with a sudden expansion stroke operates in the heat engine mode for $z\geq (\sqrt{8\tau+1}-1)/2$. For the refrigeration operational mode, the positive cooling condition $Q_{c(SE)}^{\rm HT}>0$ implies that $z^2\leq 2\tau-1$, which in turn implies that $\tau$ should be greater than 1/2. This means that heat extraction from the cold reservoir is possible only when the temperature of the cold reservoir is larger than half the value of the hot reservoir temperature. The Otto cycle under consideration operates as an accelerator for $\tau\leq z\leq (\sqrt{8\tau+1}-1)/2$. Finally, the heater mode is realized for $2\tau-1\leq z^2\leq\tau^2$. For the complete phase diagram, see left panel of Fig.~\ref{phase_diagram}.

\subsection{Asymmetric Otto cycle with sudden compression stroke}

For the sudden compression stroke, the expression for extracted work  is given in Eq. (\ref{WSCHT}).  In the high-temperature limit, the expressions for $Q_{h(SC)}^{\rm HT}$ and $Q_{c(SC)}^{\rm HT} $ can be obtained using Eqs. (\ref{heat2}) and (\ref{heat4}), and are given by
\begin{equation}
Q_{h(SC)}^{\rm HT} = \frac{1}{\beta_h}\left(1 - \frac{1+z^2}{2z^2}\tau\right), 
\quad 
Q_{c(SC)}^{\rm HT} =     \frac{1}{\beta_h} (\tau - z). \label{QhcSC}
\end{equation}
 In this case, the realization of four different modes are explained as follows: the heat engine and refrigeration modes are achieved for $z\geq (\sqrt{\tau(8+\tau)}+\tau)/4$ and $z\leq \tau$, respectively; the accelerator mode is achieved for the parametric regime $\sqrt{\tau/(2-\tau)}\leq z \leq (\sqrt{\tau(8+\tau)}+\tau)/4$; finally, the Otto cycle operates as a heater for $\tau\leq z\leq \sqrt{\tau/(2-\tau)}$. It is interesting to note that in this case, exactly half of the parametric space is available for the refrigeration operation, whereas, for the Otto cycle with sudden expansion stroke, the same parametric space was available for the refrigeration and the heater modes. Additionally, in this sudden compression stroke case, the parametric regime in ($\tau,z$) space available for both heater and accelerator modes shrinks as compared to the sudden expansion case. Once again, for the complete phase diagram, see right panel of Fig.~\ref{phase_diagram}.

\section{Conclusions}\label{sec7}

In this paper, we have studied the performance of the quantum Otto cycle with a harmonic oscillator as the working medium (viewed as a heat engine) and with one of the two work strokes being \textit{non-adiabatic}. To perform analytic computations for this model, we assumed that the non-adiabatic stroke was a \textit{sudden quench}. With this assumption, we derived analytical upper bounds for the efficiency of the engine.

By considering the high-temperature regime, we have shown that the upper bound on the efficiency is qualitatively higher when the sudden quench is in the compression rather than in the expansion stroke. The difference in these bounds is quantified by the differences between Eqs. \eqref{Taylor} and \eqref{Taylor2} (which begins from second order in the Taylor expansion). Thus, we conclude that the two working configurations are highly asymmetric in terms of their respective performances.
Moreover, we have shown that the Otto cycle cannot work as an engine in the low-temperature regime, and we have provided heuristics and numerical evidence that the analytical upper bounds obtained in the high-temperature regime can be extended to be valid in all temperature regimes. Then, by carrying out a detailed comparison of the performances of the Otto engine for three different driving protocols: sudden compression, sudden expansion, and sudden switch, we highlighted some connections among the different operational points for these driving schemes.  
Finally, we have shown that the presence of frictional effects, induced by the sudden quench, modifies the PWC of the cycle. As a result, the Otto cycle acquires non-trivial phase diagrams of its operating regimes, with the additional possibilities of being used as a heater or as a thermal accelerator, beyond the usual operating regimes of use as an engine and a refrigerator.

We believe that our results will be of relevance in the recent surge of literature on the shortcuts to adiabaticity, \cite{STAreview}. In fact, we have shown a very clear case in which two, seemingly symmetric, non-adiabatic processes have very different effects on the performance of a thermodynamic device, thus showing that the efforts towards finding shortcuts to adiabaticity should be concentrated on the expansion process rather than on the compression process. It therefore becomes an interesting question to consider the generality of our findings. For example, their applicability to: other thermodynamic cycles beyond the Otto cycle; other models of the working medium (beyond the simple harmonic oscillator); or other protocols for the expansion/compression strokes. For example, one may wonder how this asymmetry behaves when many-body systems are used as the working medium. 
Furthermore, we notice that our results are similar in spirit (in a sense, dual) to recent literature discussing the asymmetry of cooling and heating, \cite{HasegawaPRR2021,Alessio2020}, as these works show that the processes of cooling and heating are also not symmetric for thermodynamic purposes. It would be of clear interest to explore this connection further.

We hope to come back to all these intriguing open points in the near future.

\section*{Acknowledgements}
Vahid Shaghaghi, Varinder Singh, and Dario Rosa acknowledge support from the Institute for Basic Science in Korea (IBS-R024-D1). Cameron Beetar would like to thank the Isaac Newton Institute for Mathematical Sciences for support and hospitality during the programme \textit{Black holes: bridges between number theory and holographic quantum information} when work on this paper was completed; EPSRC grant number EP/R014604/1. Giuliano Benenti acknowledges financial support by:  Julian Schwinger Foundation (Grant JSF-21-04-0001), PRIN MUR (Grant No. 2022XK5CPX) and INFN through the project “QUANTUM”.

\appendix\section{CASUS IRREDUCIBILIS}
%
%In algebra, Casus irreducibilis arises while solving a cubic equation.
%The formal statement of the Casus irreducibilis 
%is that if a cubic polynomial is irreducible with rational coefficients 
%and has three real roots, then the roots of 
%the cubic equation are not expressible using real radicals and thus, 
%one must introduce expressions with complex radicals,
%even though the resulting expressions are actually real-valued.
%%
%It was proven by P. Wantzel in 1843 \cite{Kleiner2007}.  Using the 
%discriminant $D$ of the irreducible cubic equation, 
%one can decide whether the given equation is in Casus irreducibilis or not,
%via Cardano's formula \cite{Stewart1990}.
%
\begin{widetext}
While dealing with a cubic equation, the case of \textit{casus irreducibilis} \cite{Kleiner2007,Stewart1990} may arise  when 
the discriminant $D=18abcd-4b^3d+b^2c^2-4ac^3-27a^2d^2$ of the equation 
\begin{equation}
a x^3 + b x^2 + c x + d  =0, \qquad (a, b, c, d \text{\, are real}) \label{AZ1}
\end{equation}
is always positive, i.e., $D>0$.
In such a such case, even if all the roots are real, they cannot be expressed without using complex radicals. However, a solution can be obtained in terms of trigonometric functions as is explained below. Eq. (\ref{AZ1}) can be written in the following form:
\begin{equation}
    x^3 + A x^2 + B x + C = 0, \label{AZ2}
\end{equation}
where $A=b/a$, $B=c/a$ and $C=d/a$. The solution of the above equation is given by \cite{BarnettBook}
\begin{equation}
    x=   -\frac{A}{3}  +  \frac{2}{3} \sqrt{A^2-3 B} \cos \left(\frac{1}{3} \cos ^{-1}\left(-\frac{2 A^3-9 A B+27 \text{C1}}{2 \left(A^2-3 B\right)^{3/2}}\right)\right) . \label{solutionz}
\end{equation}

In our case, the discriminant $D$ of the cubic equation,
\begin{equation}
     z^3 -\frac{3}{2} z^2\tau +\frac{1}{2}\tau (2\tau-1)=0, \label{AZ3}
\end{equation}
is $D =   108 (1-\tau)^2 \tau ^2 (2\tau-1)>0$ for $\tau>1/2$, thus presenting us with a \textit{casus   irreducibilis}. Comparing Eq. (\ref{AZ3}) with Eq. (\ref{AZ2}), we identify $A=-3\tau/2$, $B=0$ and $C=\tau(2\tau-1)/2$. Further, by using Eq. (\ref{solutionz}), we obtain the solution for $z$ given in Eq. (\ref{auxA}) in the main text.

\end{widetext}
\bibliography{QHE-reference}
\bibliographystyle{apsrev4-1}

\end{document}